\documentclass[a4paper,10pt]{article}

\usepackage[dvips]{graphicx}
\usepackage{amssymb}

\newtheorem{theorem}{Theorem}[section]

\newtheorem{lemma}{Lemma}[section]

\newtheorem{remark}{Remark}[section]

\title{A unified approach to determining the early exercise boundary position at expiry for American style of general class of derivatives}
\author{Tom\'a\v{s} Bokes\\
{\normalsize Department of Applied Mathematics and Statistics,}\\ 
{\normalsize Faculty of Mathematics, Physics and Informatics,} \\
{\normalsize Comenius University, 842 48 Bratislava, Slovak Republic}\\
{\normalsize bokes@fmph.uniba.sk}}

\begin{document}

\maketitle

\begin{abstract}
	In this paper, we present a unified method for calculating the limit of early exercise boundary at expiry.
	We price American style of general derivative using a formula expressed as a sum of the value of
	European style of derivative and so called American premium. We use the latter expression to calculate
	an analytic formula for limit of early exercise boundary at expiry. Method applied on American style plain
	vanilla, Asian and lookback options yields identical results with already known values. Results for selected
	American style of derivative strategies are compared with limits calculated by the PSOR method.


\noindent {\bf AMS-MOS classification:} 91B28\hspace{1cm}{\bf JEL classification:} C65, G13

\end{abstract}

\pagestyle{myheadings}
\thispagestyle{plain}
\markboth{}{A unified approach to determining the early exercise boundary position at expiry}

\section{Introduction}
	The growth of variety in financial derivatives traded on markets has increased the need for more
	general and more accurate valuation of their prices. The breakpoint in valuation methods for the financial
	derivatives is dated to early 70. of the $20^{th}$ century. The cornerstone laid by Black, Scholes and Merton in
	Black and Scholes (1973) and Merton (1973) or its modifications occur in majority of all pricing techniques. The well known
	Black-Scholes partial differential equation	and the theory behind it are considered as an important basis
	in the financial engineering. However, the theory of valuation has undergone many changes since that time.
	
	The most basic classification of the financial derivatives is according to their expiration time (one of the main
	properties). The European style of derivatives can be exercised only at the expiration time $T$. On the other hand,
	by buying the American style of derivatives the holder obtains a right to exercise it at any moment by the expiration time.
	The early exercise boundary of financial derivative $x_t^*=x^*(t)$ splits the $t-x$ (time--underlying) space into
	continuation region $\mathcal{C}$ and stopping region $\mathcal{S}$. The derivative is exercised if spot value
	of the underlying is in the stopping region, i.e. $(t,x_t)\in\mathcal{S}$, the derivative is held otherwise.
	
	We created a unified approach for calculation of the analytic value
	of limit of the early exercise boundary at expiry. Note that position of the early exercise boundary at the expiry
	was already calculated for many particular cases of derivatives Albanese and Campolieti (2006), Alobaidi and Mallier (2006),
	Bokes and {\v S}ev{\v c}ovi{\v c} (2011), Chiarella and Ziogas (2005), Dai and Kwok (2006), Detemple (2006), Kwok (2008),
	{\v S}ev{\v c}ovi{\v c} (2008), Wilmott et al. (1995), Wu et al. (1999), etc.
	In this paper, we present theorem that generalizes these methods into one formula. It can be used to determine the limit
	of exercise boundary for American style of a general derivative that can be written in desired form (\ref{am_price_intro}).
	
	In the second section of this paper, we summarize a method of valuation of the American style of derivatives
	driven by a Brownian motion (it can be used to calculate the value of a derivative in form required by following method).
	In the third section, we present the main result of the paper. The method of calculation of
	the limit of early exercise boundary is presented in Theorem~\ref{st_p}. Next section
	consists of several examples where the method is applied on plain vanilla
	options, American style of their strategies, on the Asian and lookback options, shout options and British style of
	vanilla option. In the last section, we compare the position of early exercise boundary (of condor spread) calculated
	by our theorem and values calculated by the classical PSOR method.
	
\section{Value of the American style of derivatives}

	In this paper, we analyze the American style of derivative with price given by the formula
	\begin{equation}
		\label{am_price_intro}
		V_{am}(t,x_t)=V_{eu}(t,x_t)+\mathbb{E}_t\left[\int_t^T{{\bf1}_{\mathcal{S}}(u,x_u)f_{b}(u,x_u)}\,du\right],
	\end{equation}
	where $V_{eu}$ is the price of European style of derivative, $\mathbb{E}_t$ is conditioned expected value according
	to the information at time $t$, ${\bf1}_{\mathcal{S}}$ is the indicator function for stopping region $\mathcal{S}$
	and $f_{b}$ is American style bonus function. The value of an American style vanilla option in this form was first
	introduced by Kim (1990).
	
	\begin{remark}
		Price process of the American style derivative discounted by the numeraire is a supermartingale according to the
		risk neutral measure. It is the Snell envelope of the pay-off process discounted by the numeraire and
		(\ref{am_price_intro}) discounted by the numeraire is the Doob-Meyer decomposition of this supermartingale.
		For further details see (Karatzas and Shreve, 1988; Chapter 1).
	\end{remark}
	
	As an example, we can assume an underlying driven by a stochastic differential equation
	$$
		dx_t^i=\mu^i\ dt+\sigma^i\ dW_t^i,
	$$
	for $i\in\{1,\ldots, n\}$ on their domain $\mathbb{D}\subset\mathbb{R}^n$. The values $\mu^i\in\mathbb{R}$,
	$\sigma^i\geq0$ and $dW_t=(dW_t^1, \ldots, dW_t^n)$ are drift, volatility and differential of standard $n$-dimensional
	Brownian motion under the joined risk-neutral measure $\mathcal{Q}$, respectively. The covariance matrix
	of $dW_t$ is defined for $i,j\in\{1, \ldots, n\}$ by
	$$
		\mathbb{C}ovar\left[dW_t^i,dW_t^j\right]=\rho_{ij} dt,
	$$
	where $\rho_{ij}\in[-1,1]$ is the correlation coefficient and $\rho_{i,i}=1$.
	
	Let $\Omega$ and $\mathcal{N}$ be the pay-off function and the numeraire, respectively. The value $V(t,x_t)$ of an
	American style of derivative on the underlying asset $x_t$ is then given by
	$$
		V(t,x_t)=v(t,x_t)+e(t,x_t),
	$$
	where
	\begin{eqnarray*}
		v(t,x_t)&\equiv &\mathcal{N}(t, x_t)\ \mathbb{E}_t^\mathcal{Q}\left[\left(\mathcal{N}(T, x_T)\right)^{-1}\Omega(T, x_T)\right],\\
		e(t,x_t)&\equiv &\mathcal{N}(t, x_t)\ \mathbb{E}_t^\mathcal{Q}\left[-\int_t^T{{\bf1}_{\mathcal{S}}(u,x_u) f_d(u,x_u)\,du}\right],
	\end{eqnarray*}
	and
	$$
		f_d(t,x_t)=\frac{\partial\left(\frac{\Omega(t, x_t)}{\mathcal{N}(t, x_t)}\right)}{\partial t}
						+\sum_{i=1}^{n}{\mu^i\frac{\partial\left(\frac{\Omega(t, x_t)}{\mathcal{N}(t, x_t)}\right)}{\partial x^i}}
						+\frac12\sum_{i,j=1}^{n}{\rho_{ij}\sigma^i\sigma^j\frac{\partial^2\left(\frac{\Omega(t, x_t)}{\mathcal{N}(t, x_t)}\right)}{\partial x^i\partial x^j}}.
	$$
	
	The bonus function in (\ref{am_price_intro}) at expiry is then given by the expression
	\begin{equation}
		\label{bonus_deriv}
		f_{b}(T,x_T)=-\mathcal{N}(T, x_T)f_d(T,x_T).
	\end{equation}
	Values of $f_d$ on the set of zero measure where $f$ and $\mathcal{N}$ are not differentiable are defined as
	the arithmetic average of limes superior and limes inferior at each point of this set.

\section{Limit of the early exercise boundary at expiry}

	In this section, we determine the main result of the paper - the position of the early exercise boundary
	$x^*_T$ at expiry $T$ for a general class of financial derivatives. The result is stated for a wide class of integral
	equations for pricing American style of derivatives of the form (\ref{gen_form}). This problem has been already considered
	by many authors for American style of certain derivatives (e.g. Albanese and Campolieti, 2006; Alobaidi and Mallier, 2006;
	Bokes and {\v S}ev{\v c}ovi{\v c}, 2011; Chiarella and Ziogas, 2005; Dai and Kwok, 2006; Detemple, 2006; Kwok, 2008; {\v S}ev{\v c}ovi{\v c}, 2008; Wilmott et al., 1995; Wu et al., 1999).
	Presented method is a unified approach solving the generalized
	problem of finding the position of the early exercise boundary at expiry.
	
	Let $\mathbb{D}\subset\mathbb{R}^n$ be a subset of Euclidean space $\mathbb{R}^n$.
	In what follows, we shall denote by $\partial A$ a boundary of the set $A\subset\mathbb{D}$ with respect to the
	topology $\mathbb{D}$, i.e. $\partial A=\overline{A}\cap\overline{\mathbb{D}\backslash A}$.

	\begin{theorem}
		\label{st_p}
		Consider an American style of derivative $V_{am}$ on the underlying $x\in\mathbb{D}\subset\mathbb{R}^n$ with
		the stopping and continuation regions defined by the open sets $\mathcal{S}\subset\mathbb{D}$ and
		$\mathcal{C}\subset\mathbb{D}$, respectively. Let $\mathcal{X}^*_t=\partial\mathcal{S}(t,.)\equiv \partial\mathcal{C}(t,.)$  for $t\in[0,T]$ be
		a (set of) manifold(s) of the early exercise boundary at time $t$. Suppose that the value of $V_{am}$ is given
		by the equation 
		\begin{equation}
			\label{gen_form}
			V_{am}(t,x_t)=V_{eu}(t,x_t)+\mathbb{E}_t\left[\int_t^T{{\bf1}_{\mathcal{S}}(u,x_u)f_{b}(u,x_u)}\,du\right],
		\end{equation}
		where $V_{eu}$ denotes a price of the corresponding European style of derivative and $f_{b}(t,x)$ is
		a function representing the early exercise bonus. Furthermore, we suppose that
		\begin{equation}
			\label{h1}
			V_{am}(t,x)\geq \Omega(t,x) \hbox{ and } V_{am}(t,x)\geq V_{eu}(t,x) \quad \hbox{for any}\ t\in [0,T], x\in\mathbb{D},
		\end{equation}
		where $\Omega(t,x)$ is the pay-off function at time $t$ for both American style and European style of
		derivative, i.e.
		\begin{equation}
			\label{h2}
			V_{am}(T,x)=\Omega(T,x)=V_{eu}(T,x) \quad \hbox{for any}\ x\in\mathbb{D}.
		\end{equation}
		Then the limit of early exercise boundary at expiry is given by
		\begin{equation}
			\label{st_point_def}
			\mathcal{X}^*_T=\partial Z_T^+,
		\end{equation}
		where $Z_T^+=\{x_T\in\mathbb{D}; f_{b}(T,x_T)>0\}$.
	\end{theorem} 
	
	\begin{lemma}
		\label{sets}
		Consider a mutually disjoint decomposition $\mathbb{D}=A\cup\partial A\cup B$ of a topological space $\mathbb{D}$,
		where $\partial A\equiv\partial B$. Moreover, consider a set
		$Z$ so that $A\subset Z \subset \overline{A}$, where $\overline{A} \equiv A\cup\partial A$ is the closure of the set
		$A$, then $\partial A=\partial Z$.
	\end{lemma}
	
	\medskip
	\noindent P r o o f: [{\rm of Lemma~\ref{sets}}]
	
		Let $a\in\partial A$. For each $\varepsilon>0$, there exists a neighborhood $O_{\varepsilon}(a)$ so
		that $\widetilde{a}\in A\cap O_{\varepsilon}(a)$ and $\widetilde{b}\in B\cap O_{\varepsilon}(a)$.
		This implies, that $a\in\partial Z$, i.e.
		$\partial A\subset\partial Z$, because $\widetilde{a}\in Z$, but
		$\widetilde{b}\not\in Z\subset \overline{A}=A\cup\partial A$.
	
		Since $\partial A=\overline{A}\cap\overline{\mathbb{D}\backslash A}$ and $\overline{Z}=\overline{A}$, we have
		$\partial Z=\overline{Z}\cap\overline{\mathbb{D}\backslash Z}\subset\partial A$ and the proof of Lemma follows.
	\hfill$\square$\medskip

	\medskip
	\noindent P r o o f: [{\rm of Theorem~\ref{st_p}}]
		
		Part 1). First, we show that
		$$
			\mathcal{S}(T,.)\subset\{x_T\in\mathbb{D}; f_{b}(T,x_T)>0\}.
		$$
		We have
		$$
			\frac{1}{T-t}\mathbb{E}_t\left[\int_t^T{{\bf1}_{\mathcal{S}}(u,x_u)f_{b}(u,x_u)}\,du\right]
			= \frac{1}{T-t}\left( V_{am}(t,x_t)-V_{eu}(t,x_t) \right) \geq 0,
		$$
		for any $t\in [0,T)$ by (\ref{h1}). In the limit $t\to T$, we can omit the conditioned expected value operator
		$\mathbb{E}_t$ and we obtain
		$$
			{\bf1}_{\mathcal{S}}(T,x_T)f_{b}(T,x_T)\geq0. 
		$$
		If $(T,y_T) \in S$, then we obtain 
		$$
			f_{b}(T,y_T)\geq 0.
		$$
		Now suppose that there exists $(T,\widetilde{y}_T) \in \mathcal{S}$ such that $f_{b}(T,\widetilde{y}_T)= 0$.
		Notice that in the stopping region $\mathcal{S}$ we have the identity $V_{am}(t,x) = \Omega(t,x)$
		for any $(t,x)\in \mathcal{S}$ and, consequently,
		$\frac{\partial}{\partial t}\left(V_{am}(t,x)-\Omega(t,x)\right)=0$.
		Then we have
		\begin{eqnarray*}
			0=\frac{\partial}{\partial t}\left(V_{am}(T,\widetilde{y}_T)-\Omega(T,\widetilde{y}_T)\right)&=&
			\frac{\partial V_{eu}}{\partial t} (T,\widetilde{y}_T) - f_{b}(T,\widetilde{y}_T)
			-\frac{\partial \Omega}{\partial t} (T,\widetilde{y}_T)\\
			&=&
			\frac{\partial}{\partial t}\left(V_{eu}(T,\widetilde{y}_T)-\Omega(T,\widetilde{y}_T)\right).
		\end{eqnarray*}
		In the stopping region, exercising the derivative (American style) gives holder higher pay-off than
		keeping it (European style), i.e.
		$$
			V_{eu}(t,y)<\Omega(t,y) \quad \hbox{for } (t,y)\in\mathcal{S},
		$$
		for $t$ sufficiently close to expiry $T$. According to (\ref{h2}), the value of difference between European style of derivative and pay-off
		is increasing (from negative values to zero at maturity). The derivative of this difference is positive in the
		stopping region. This is a contradiction and the proof of first part follows.
		
		\medskip
		Part 2). Now, we show that
		$$
			\{x_T\in\mathbb{D}; f_{b}(T,x_T)>0\}\subset\overline{\mathcal{S}}(T,.)=\mathcal{S}(T,.)\cup\mathcal{X}^*_T.
		$$
		The function $f_{b}$ can be determined on the stopping region by the following property
		\begin{eqnarray*}
		0&=&\frac{\partial}{\partial t}\left(V_{am}(T,\widetilde{y}_T)-\Omega(T,\widetilde{y}_T)\right)\\
			&=&\frac{\partial V_{eu}}{\partial t} (T,\widetilde{y}_T) - {\bf1}_{\mathcal{S}}(T,\widetilde{y}_T)f_{b}(T,\widetilde{y}_T)
			-\frac{\partial \Omega}{\partial t} (T,\widetilde{y}_T).
		\end{eqnarray*}
		To span the function on the whole domain $\mathbb{D}$, we omit the function ${\bf1}_{\mathcal{S}}(u,x_u)$ and we have
		$$
		f_{b}(T,\widetilde{y}_T)=
		\frac{\partial}{\partial t}\left(V_{eu}(T,\widetilde{y}_T)-\Omega(T,\widetilde{y}_T)\right).
		$$
		Notice, that the function $f_{b}$ nullifies movements from the pay-off function $\Omega$.
		
		On the continuous region $\mathcal{C}$, the holder of a financial derivative does not want to exercise it,
		because keeping this derivative yields better pay-off, i.e.
		$$
			V_{eu}(t,y)>\Omega(t,y) \quad \hbox{for } (t,y)\in\mathcal{C},
		$$
		for $t$ sufficiently close to expiry $T$. According to (\ref{h2}), the value of difference between European style
		of derivative and pay-off is decreasing (from positive values to zero at maturity). The derivative of this difference
		is negative and so is the value of function $f_{b}$ in the continuation region. The function $f_{b}$ have positive
		values only on $\overline{\mathcal{S}}(T,.)$.
		
		Using Lemma~\ref{sets} we have (\ref{st_point_def}) and the proof follows.
	\hfill$\square$\medskip

	\begin{remark}
		\label{st_p_remark}
		Notice, that according to the second part of the proof of Theorem~\ref{st_p}, we can
		determine function of the American style bonus function $f_{b}$ at expiry by the formula
		$$
			f_{b}(T,y)=\lim_{t\rightarrow T}
		\frac{\partial}{\partial t}\left(V_{eu}(t,y)-\Omega(t,y)\right).
		$$
	\end{remark}
	
	\begin{remark}
		The limit of early exercise boundary analyzed in this paper is the expansion of the zeroth order.
		For several financial derivatives of American style, higher order expansion was already calculated.
		Further details on this expansion can be found in Dewynne et al. (1993), {\v S}ev{\v c}ovi{\v c} (2001), Wilmott et al. (1995)
		for plain vanilla the call option, in Stamicar et al. (1999), Zhu (2006), Zhu and He (2007) for the plain vanilla put
		option and in Bokes and {\v S}ev{\v c}ovi{\v c} (2011) for average strike Asian options.
	\end{remark}
	
\section{Calculation of the early exercise boundary position at expiry}
\label{calc}

	In this section, we calculate the limit of early exercise boundary at expiry for several types of American style of
	financial derivatives and their strategies. The underlying of all derivatives presented in this section
	is driven by a geometric Brownian motion. Theorem~\ref{st_p} does not have limitation on
	the distribution of underlying and can be used also in other models for underlying assets. The method can
	be used also e.g. on L\'evy processes - the most simple approach is to apply Remark~\ref{st_p_remark} (either analytically
	or numerically).
	
	We use parameters $r$, $q$ and $\sigma$ to denote risk-free continuous interest rate, continuous dividend
	rate and volatility of underlying, respectively. The underlying asset $S$ is driven by stochastic differential
	equation
	\begin{equation}
		\label{stdif_s}
		dS_t=(r-q)S_t dt+\sigma S_t dW_t,
	\end{equation}
	where $W_t$ is the Wiener process.
	
	We present examples of application of presented method. Results for the most basic
	of them are well known (e.g. vanilla options or Asian options with arithmetic and geometric averaging).
	The position of early exercise boundary for many other derivatives is calculated by complicated methods or
	specified only by the argumentation without any mathematical formulation. However, we include the derivation for
	comparison purposes.
	
	Some of the examples presented are American types of strategies of vanilla options. Trading these derivatives
	is not very common, we use them only to demonstrate Theorem~\ref{st_p} on more complex types of derivatives.
	
	\subsection{Plain vanilla options}
	
\begin{figure}[pt]
			\begin{center}
				\includegraphics[width=4.5cm]{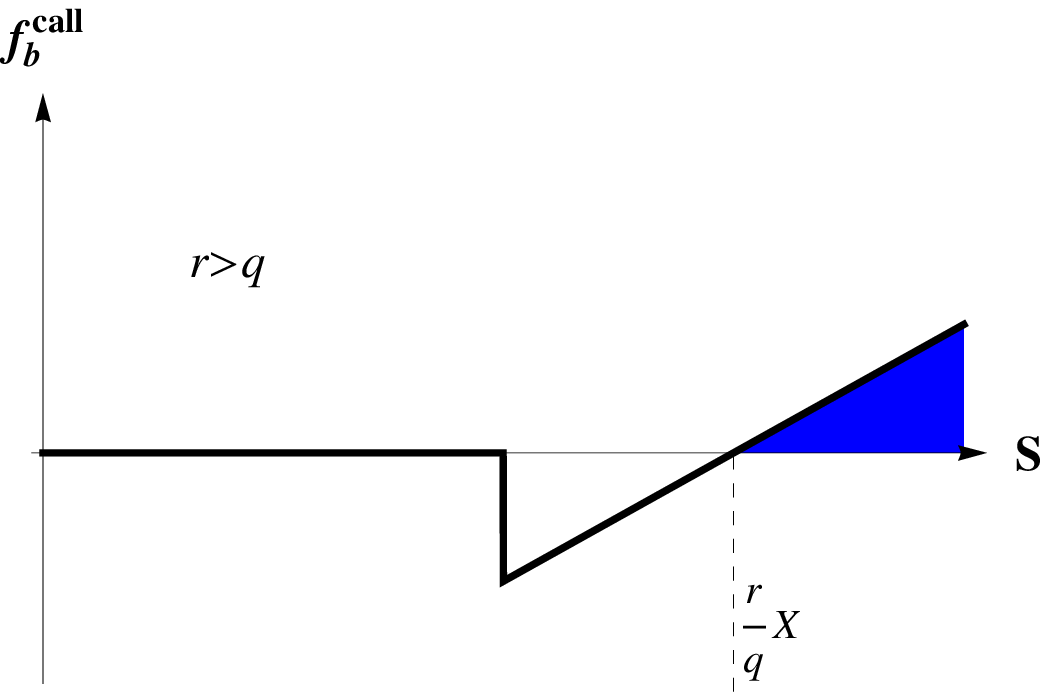}
				\includegraphics[width=4.5cm]{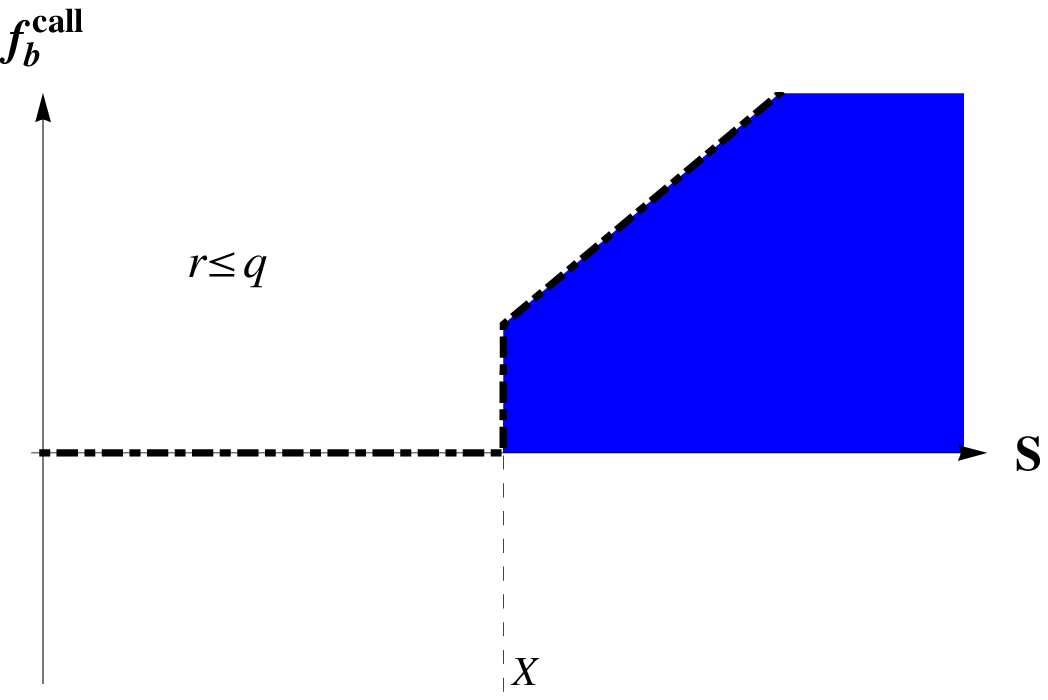}
			\end{center}
			\vspace*{8pt}
			\caption{The American style bonus function for a call option with $r>q$ (left)
						and $r\leq q$ (right).}
			\label{vanilla_call}
		\end{figure}

\begin{figure}[pt]
			\begin{center}
				\includegraphics[width=4.5cm]{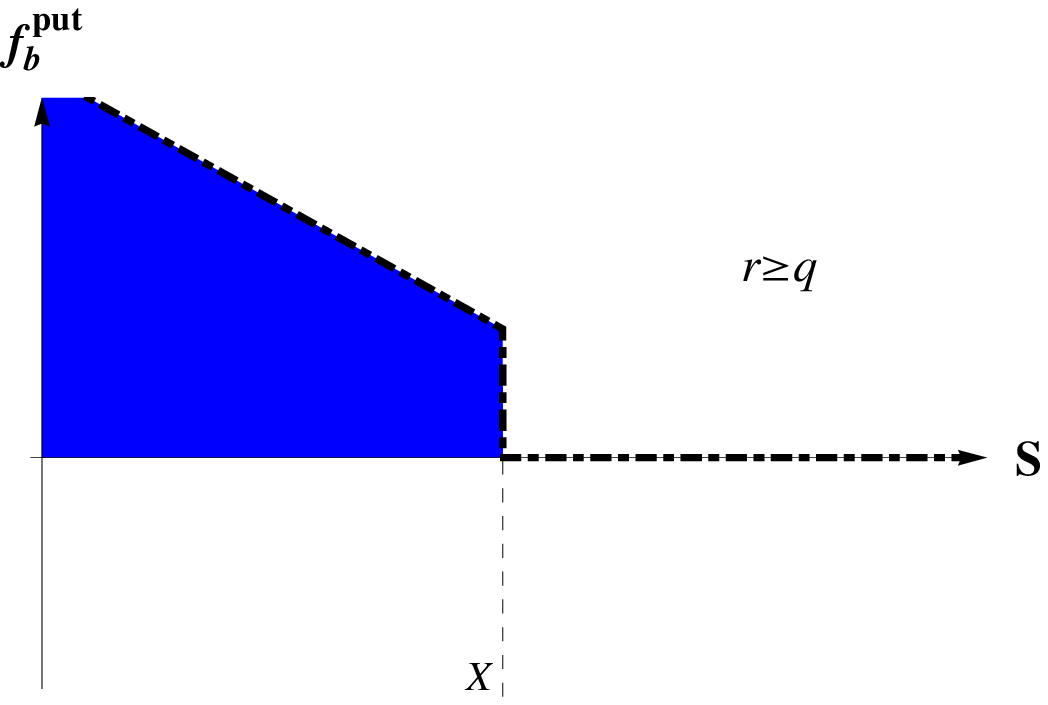}
				\includegraphics[width=4.5cm]{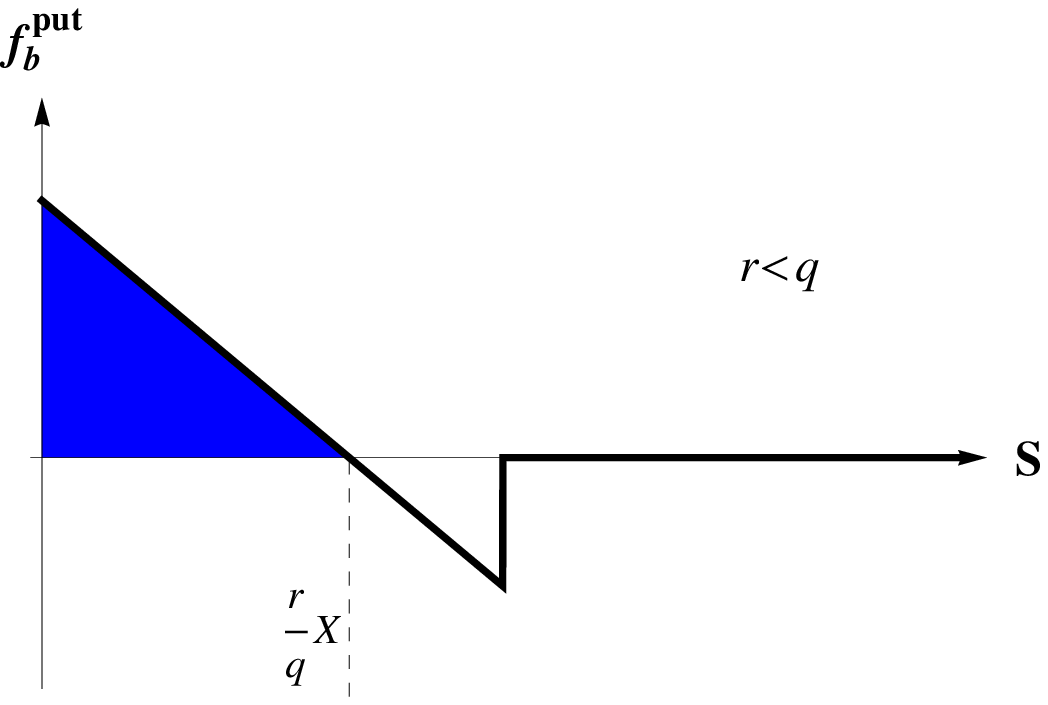}
			\end{center}
			\vspace*{8pt}
			\caption{The American style bonus function for a put option with $r>q$ (left)
						and $r\leq q$ (right).}
			\label{vanilla_put}
		\end{figure}

		The European style of vanilla call/put option gives its holder right to buy/sell the underlying $S$ at maturity
		time $T$ for the expiration price $X$.
		The position of the early exercise boundary at expiry for call and put vanilla option is equal to boundary of set
		of positive values of $f^{call}_{b}$ (see Figure~\ref{vanilla_call}) and $f^{put}_{b}$ (see Figure~\ref{vanilla_put}), i.e.
		$$
			\partial Z^{+call}_T=\max\left[X,\frac{r}{q}X\right]=S^{*call}_T
		\textrm{ and }
			\partial Z^{+put}_T=\min\left[X,\frac{r}{q}X\right]=S^{*put}_T,
		$$
		respectively. This result is well known and can be found also in Albanese and Campolieti (2006), Detemple (2006), Kwok (2008),
		Wilmott et al. (1995) and many other sources.
		
		\smallskip
		We derive the position of the early exercise boundary according to the method presented in this paper.
		The pay-off functions for call and put options are
		$$
			\Omega^{call}(t,S;X)=\left(S-X\right)^+
			\textrm{ and }
			\Omega^{put}(t,S;X)=\left(X-S\right)^+,
		$$
		respectively. The value of European style of vanilla option (the well known solution of Black-Scholes
		partial differential equation extended by Merton) for both call and put option is given by
		\begin{eqnarray}
			\label{call_value}
			C_{eu}(t,S;X)&=&e^{-q(T-t)}S\Phi\left(d_t\right)
					-e^{-r(T-t)}X\Phi\left(d_t-\sigma\sqrt{T-t}\right),\\
			\label{put_value}
			P_{eu}(t,S;X)&=&e^{-r(T-t)}X\Phi\left(-d_t+\sigma\sqrt{T-t}\right)
					-e^{-q(T-t)}S\Phi\left(-d_t\right),
		\end{eqnarray}
		where $d_t=\frac{\ln{\frac{S}{X}}+\left(r-q+\frac{\sigma^2}{2}\right)(T-t)}{\sigma\sqrt{T-t}}$.

		We know the value of both pay-off function and European style of option, so we can calculate $f_{b}$
		at the expiry according to Remark~\ref{st_p_remark}, i.e. for call option and put option we have
		$$
			f^{call}_{b}(T,S)=\left\{\begin{array}{ll}0&\textrm{for }S<X,\\
														\frac{X}2 \left(q-r\right)&\textrm{for }S=X,\\
														q S-r X&\textrm{for }S>X
										\end{array}\right.
		\textrm{ and }
			f^{put}_{b}(T,S)=\left\{\begin{array}{ll}r X-q S&\textrm{for }S<X,\\
														\frac{X}2 \left(r-q\right)&\textrm{for }S=X,\\
														0&\textrm{for }S>X,
										\end{array}\right.
		$$
		respectively.
	
	\subsection{Option strategies}
	
		There are many option strategies that can be considered. European style of these financial derivatives is linear
		combination of plain vanilla options, therefore the derivation is similar for all of them. The position of early
		exercise boundary at expiry is calculated only for very few of them. The results for American style strangle
		spread was presented by Alobaidi and Mallier (2006), Chiarella and Ziogas (2005).
		
		The most complex frequently used strategy consisting of vanilla options is European style of condor spread
		and its restriction butterfly spread. The European style of condor spread is a linear combination of four
		vanilla call options. There are three different cases, when determining the position of the early exercise boundary
		at expiry by the boundary of set of positive values of $f^{condor}_{b}$.
		
\begin{figure}[pt]
			\begin{center}
				\includegraphics[width=4cm]{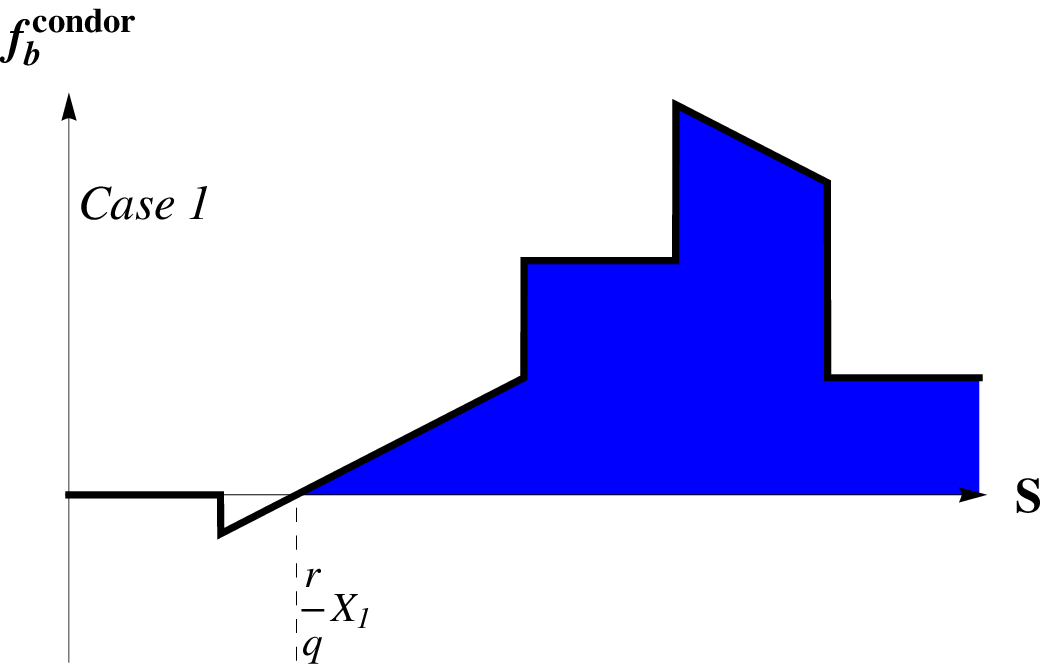}
				\includegraphics[width=4cm]{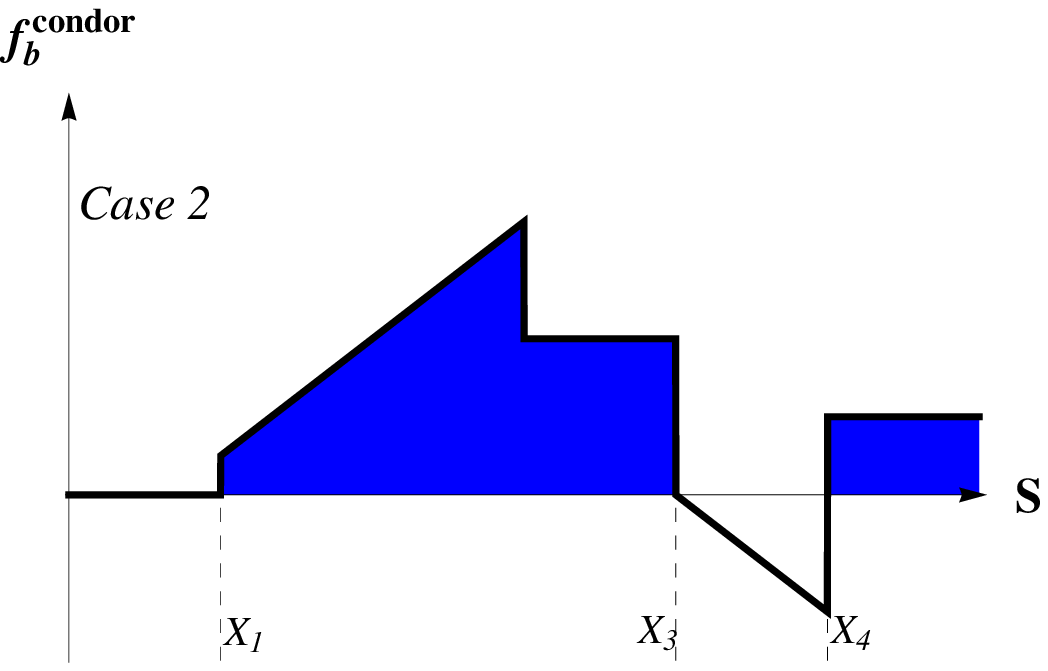}
				\includegraphics[width=4cm]{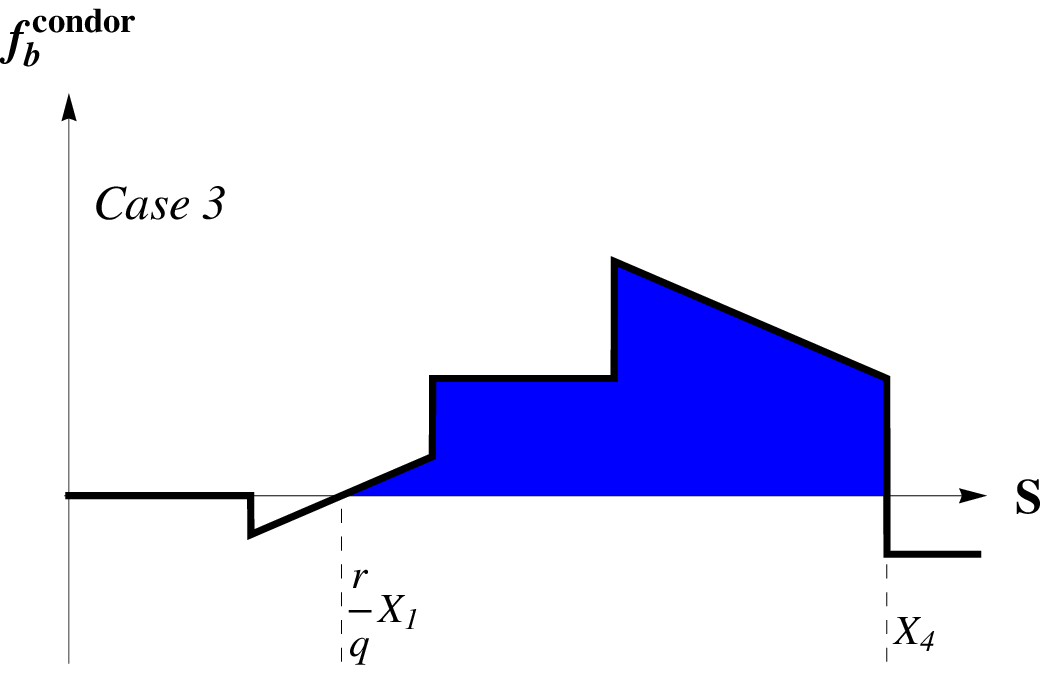}
			\end{center}
			\vspace*{8pt}
			\caption{The American style bonus function for condor spread with $-X_4+X_3+X_2-X_1>0$ and
						$r\left(X_3+X_2-X_1\right)\geq q X_4$ (left), $-X_4+X_3+X_2-X_1>0$ and
						$r\left(X_3+X_2-X_1\right)< q X_4$ (center) and $-X_4+X_3+X_2-X_1\leq 0$ (right)}
			\label{condor}
		\end{figure}

		In the first case, if we have $-X_4+X_3+X_2-X_1>0$ and $r\left(X_3+X_2-X_1\right)\geq q X_4$, then the set 
		of boundary points has only one element (see Figure~\ref{condor} (left)).
		$$
			\partial Z^{+condor}_T=\min\left[\max\left[X_1,\frac{r}{q}X_1\right],X_2\right]=S^{*condor}_T.
		$$
		
		In the second case, we have $-X_4+X_3+X_2-X_1>0$ and $r\left(X_3+X_2-X_1\right)< q X_4$. The set of boundary points has
		three elements (see Figure~\ref{condor} (center)).
		\begin{eqnarray*}
			\partial Z^{+condor}_T&=&\left\{\min\left[\max\left[X_1,\frac{r}{q}X_1\right],X_2\right],\right.\\
									&&\quad\left.\max\left[X_3,\frac{r}{q}\left(X_3+X_2-X_1\right)\right],X_4\right\}=S^{*condor}_T.
		\end{eqnarray*}
		
		In the last case, we have $-X_4+X_3+X_2-X_1\leq 0$ and the set of boundary points has two elements
		(see Figure~\ref{condor} (right)).
		\begin{eqnarray*}
			\partial Z^{+condor}_T&=&\left\{\min\left[\max\left[X_1,\frac{r}{q}X_1\right],X_2\right],\right.\\
										&&\quad\left.\min\left[\max\left[X_3,\frac{r}{q}\left(X_3+X_2-X_1\right)\right],X_4\right]\right\}=S^{*condor}_T.
		\end{eqnarray*}
	
		\smallskip
		We present derivation for the position of the early exercise boundary at expiry for a condor spread according to
		the method presented in this paper. The pay-off function of a condor spread is
		\begin{eqnarray*}
			\Omega^{condor}(t,S;X_1,X_2,X_3,X_4)&=&\Omega^{call}(t,S;X_1)-\Omega^{call}(t,S;X_2)\\
					&&\quad-\Omega^{call}(t,S;X_3)+\Omega^{call}(t,S;X_4)\\
				&=&\left(S-X_1\right)^+ - \left(S-X_2\right)^+ -\left(S-X_3\right)^+ + \left(S-X_4\right)^+,
		\end{eqnarray*}
		for $X_1<X_2\leq X_3<X_4$, where the case $X_2=X_3$ is called a butterfly spread. The price of a
		European style of condor is calculated by the formula
		\begin{eqnarray*}
			V^{condor}_{eu}(t,S;X_1,X_2,X_3,X_4)&=&C_{eu}(t,S;X_1)-C_{eu}(t,S;X_2)\\
						&&\quad-C_{eu}(t,S;X_3)+C_{eu}(t,S;X_4),
		\end{eqnarray*}
		where the function $C_{eu}$ is defined by (\ref{call_value}).
		
		Once more, we use Remark~\ref{st_p_remark} to calculate the bonus function for American style of
		condor spread.
		$$
			f^{condor}_{b}(T,S)=\left\{\begin{array}{ll}0&\textrm{for }S<X_1,\\
														\frac{X_1}2 \left(q-r\right)&\textrm{for }S=X_1,\\
														q S - r X_1&\textrm{for }X_1<S<X_2,\\
														\frac{X_2}2 \left(q+r\right)-r X_1&\textrm{for }S=X_2,\\
														r\left(X_2-X_1\right)&\textrm{for }X_2<S<X_3\textrm{ or }X_2=S=X_3,\\
														\frac{X_3}2 \left(r-q\right)+r\left(X_2-X_1\right)&\textrm{for }S=X_3,\\
														r\left(X_3+X_2-X_1\right) - q S&\textrm{for }X_3<S<X_4,\\
														r\left(X_3+X_2-X_1\right)-\frac{X_4}2 \left(q+r\right)&\textrm{for }S=X_4,\\
														r\left(-X_4+X_3+X_2-X_1\right)&\textrm{for }X_4<S.
										\end{array}\right.
		$$
				
	\subsection{Asian and lookback options}
	
		Options with some extra properties are usually called exotic options. Subgroup of exotic
		options called path-dependent options contains also Asian and lookback options. Asian and lookback
		options depends on some type of average ($A$) and the extreme value (maximum $M$ or minimum $m$) of the
		underlying through the lifespan of the option, respectively.
		
		We calculate the value of limit at the expiry for American style of floating strike of both Asian and lookback options.
		The pay-off functions of floating strike Asian call, Asian put, lookback call and lookback put options are
		$$
			\Omega^{Asian}_{call}(t,S,A)=\left(S-A\right)^+,
			\Omega^{Asian}_{put}(t,S,A)=\left(A-S\right)^+,
		$$
		$$
			\Omega^{lookback}_{call}(t,S,m)=\left(S-m\right)^+
		\textrm{ and }
			\Omega^{lookback}_{put}(t,S,M)=\left(M-S\right)^+,
		$$
		respectively.
		Now, we use (\ref{bonus_deriv}) to determine the bonus function $f_{b}$. According to
		Hansen and Jorgensen (2000) we can use the numeraire $\mathcal{N}_t=e^{r t}$. We define the general exponentially weighted
		average $A$ at time $t$ by the formula
		$$
			(A_t)^p=\frac{\lambda}{1-e^{-\lambda t}}\int_0^t{e^{-\lambda(t-u)}(S_u)^p}\,du,
		$$
		where $\lambda=0$ for regular and $\lambda>0$ for weighted average. By setting the parameter $p$ to values $0$,
		$1$, $-\infty$ and $\infty$, we can construct expressions for continuous geometric average, continuous arithmetic
		average, minimum value and maximum value, respectively.
		
		The function $f_d$ for call option and for put option has form
		$$
			f_d^{call}(t,S,A)=\left\{\begin{array}{ll}0&\textrm{for }S<A,\\
												\frac12\left(\limsup_{S\rightarrow A}f_d^{call}(t,S,A)+\liminf_{S\rightarrow A}f_d^{call}(t,S,A)\right)&\textrm{for }S=A,\\
												e^{-r t}\left(-r\left(S-A\right)+\mu_S-\mu_A\right)&\textrm{for }A<S
										\end{array}\right.
		$$
		and
		$$
			f_d^{put}(t,S,A)=\left\{\begin{array}{ll}e^{-r t}\left(-r\left(A-S\right)+\mu_A-\mu_S\right)&\textrm{for }S<A,\\
												\frac12\left(\limsup_{S\rightarrow A}f_d^{put}(t,S,A)+\liminf_{S\rightarrow A}f_d^{put}(t,S,A)\right)&\textrm{for }S=A,\\
												0&\textrm{for }A<S,
										\end{array}\right.
		$$
		respectively. The value $\mu_S=(r-q)S$ according to (\ref{stdif_s}) and $\mu_A$ is drift of the stochastic
		differential equation
		$$
			dA=\mu_A dt+\sigma_A dW_t^A.
		$$
		According to Hansen and Jorgensen (2000) and Bokes and {\v S}ev{\v c}ovi{\v c} (2011) we have the form of stochastic differential
		equation for geometric average
		$$
			dA^g_t=-\frac1t A^g_t\ln{\frac{A^g_t}{S_t}}\,dt,
		$$
		for the arithmetic average
		$$
			dA^a_t=\frac1t\left(S_t-A^a_t\right)\,dt,
		$$
		and for both maximum and minimum the equation has form
		$$
			dA^{-\infty}_t=dA^{\infty}_t=0.
		$$
		Finally, the stochastic differential equation for the exponentially weighted general average has form
		$$
			dA_t=A_t\frac{\lambda}{p(1-e^{-\lambda t})}\left(\left(\frac{S_t}{A_t}\right)^p-1\right)\,dt.
		$$
		The bonus function at the expiry for call and put Asian options with continuous geometric average has form
		$$
			f^{call,g}_{b}(T,S,A)=\left\{\begin{array}{ll}0&\textrm{for }S<A,\\
												\frac{A}2\left(q-r\right)&\textrm{for }S=A,\\
												-rA+qS-\frac1T A\ln{\frac{A}{S}}&\textrm{for }A<S
										\end{array}\right.
		$$
		and
		$$
			f^{put,g}_{b}(T,S,A)=\left\{\begin{array}{ll}rA-qS+\frac1T A\ln{\frac{A}{S}}&\textrm{for }S<A,\\
												\frac{A}2\left(r-q\right)&\textrm{for }S=A,\\
												0&\textrm{for }A<S,
										\end{array}\right.
		$$
		respectively. The boundary of set of positive values is given by
		$$
			\partial Z^{+call,g}_T=\left\{\left(S,A\right)\in\mathbb{R}^2_+;\frac{S}{A}=\max\left[1,\widetilde{G}\right]\right\}=\mathcal{X}^{*call,g}_T
		$$
		and
		$$
			\partial Z^{+put,g}_T=\left\{\left(S,A\right)\in\mathbb{R}^2_+;\frac{S}{A}=\min\left[1,\widetilde{G}\right]\right\}=\mathcal{X}^{*put,g}_T,
		$$
		where $\widetilde{G}$ is the positive solution of transcendental equation
		$$
			r-q G-\frac1T\ln{G}=0.
		$$
		The solution $\widetilde{G}$ is unique on $\mathbb{R}_+$ for $q\geq0$ and $T>0$.
		
		The bonus function at the expiry for call and put Asian options with continuous arithmetic average has form
		$$
			f^{call,a}_{b}(T,S,A)=\left\{\begin{array}{ll}0&\textrm{for }S<A,\\
												\frac{A}2\left(q-r\right)&\textrm{for }S=A,\\
												\left(q+\frac1T\right)S-\left(r+\frac1T\right)A&\textrm{for }A<S
										\end{array}\right.
		$$
		and
		$$
			f^{put,a}_{b}(T,S,A)=\left\{\begin{array}{ll}-\left(q+\frac1T\right)S+\left(r+\frac1T\right)A&\textrm{for }S<A,\\
												\frac{A}2\left(r-q\right)&\textrm{for }S=A,\\
												0&\textrm{for }A<S,
										\end{array}\right.
		$$
		respectively. The boundary of set of positive values is given by
		$$
			\partial Z^{+call,a}_T=\left\{\left(S,A\right)\in\mathbb{R}^2_+;\frac{S}{A}=\max\left[1,\frac{r+\frac1T}{q+\frac1T}\right]\right\}=\mathcal{X}^{*call,a}_T
		$$
		and
		$$
			\partial Z^{+put,a}_T=\left\{\left(S,A\right)\in\mathbb{R}^2_+;\frac{S}{A}=\min\left[1,\frac{r+\frac1T}{q+\frac1T}\right]\right\}=\mathcal{X}^{*put,a}_T.
		$$
		
		Finally, the bonus function at the expiry for call and put lookback options has form
		$$
			f^{min}_{b}(T,S,m)=\left\{\begin{array}{ll}0&\textrm{for }S<m,\\
												\frac{m}2\left(q-r\right)&\textrm{for }S=m,\\
												-rm+qS&\textrm{for }m<S
										\end{array}\right.
		$$
		and
		$$
			f^{max}_{b}(T,S,M)=\left\{\begin{array}{ll}rM-qS&\textrm{for }S<M,\\
												\frac{M}2\left(r-q\right)&\textrm{for }S=M,\\
												0&\textrm{for }M<S,
										\end{array}\right.
		$$
		respectively. The boundary of set of positive values is given by
		$$
			\partial Z^{+min}_T=\left\{\left(S,m\right)\in\mathbb{R}^2_+;\frac{S}{m}=\max\left[1,\frac{r}{q}\right]\right\}=\mathcal{X}^{*min}_T
		$$
		and
		$$
			\partial Z^{+max}_T=\left\{\left(S,M\right)\in\mathbb{R}^2_+;\frac{S}{M}=\min\left[1,\frac{r}{q}\right]\right\}=\mathcal{X}^{*max}_T.
		$$
		
		The same results for the limit of early exercise boundary at expiry can be found in Detemple (2006),
		Wu et al. (1999) for geometric average Asian options, in {\v S}ev{\v c}ovi{\v c} (2008) for arithmetic
		average Asian options, in Dai and Kwok (2006) for arithmetic average Asian and lookback options and in
		Bokes and {\v S}ev{\v c}ovi{\v c} (2011) for both geometric and arithmetic Asian options.
		
		This is the first time that the position of early exercise boundary at expiry for general average is mentioned
		in the literature. It can be derived similarly as the previous examples. The bonus function at the expiry for
		call and put Asian options with general exponentially weighted average has form
		$$
			f^{call,w}_{b}(T,S,A)=\left\{\begin{array}{ll}0&\textrm{for }S<A,\\
												\frac{A}2\left(q-r\right)&\textrm{for }S=A,\\
												-rA+qS+A\frac{\lambda}{p(1-e^{-\lambda T})}\left(\left(\frac{S}{A}\right)^p-1\right)&\textrm{for }A<S
										\end{array}\right.
		$$
		and
		$$
			f^{put,w}_{b}(T,S,A)=\left\{\begin{array}{ll}rA-qS-A\frac{\lambda}{p(1-e^{-\lambda T})}\left(\left(\frac{S}{A}\right)^p-1\right)&\textrm{for }S<A,\\
												\frac{A}2\left(r-q\right)&\textrm{for }S=A,\\
												0&\textrm{for }A<S,
										\end{array}\right.
		$$
		respectively. The boundary of set of positive values is given by
		$$
			\partial Z^{+call,w}_T=\left\{\left(S,A\right)\in\mathbb{R}^2_+;\frac{S}{A}=\max\left[1,\widetilde{Y}\right]\right\}=\mathcal{X}^{*call,w}_T
		$$
		and
		$$
			\partial Z^{+put,w}_T=\left\{\left(S,A\right)\in\mathbb{R}^2_+;\frac{S}{A}=\min\left[1,\widetilde{Y}\right]\right\}=\mathcal{X}^{*put,w}_T,
		$$
		where $\widetilde{Y}$ is the positive solution of transcendental equation
		$$
			r-q Y-\frac{\lambda}{p(1-e^{-\lambda T})}\left(Y^p-1\right)=0.
		$$
		The solution $\widetilde{Y}$ is unique on $\mathbb{R}_+$ for $q\geq0$ and $T>0$.
		
	\subsection{Shout options}
	
			Shout options are financial derivatives similar to European plain vanilla options. The difference is that
			the holder of a shout option can once during the life of derivative ''shout'' to the writer, i.e. the option
			expires and the strike price is reset to actual spot price of the underlying asset. The shouting action is
			conditioned by in-the-money position of the option. According to this property, we need to know optimal shouting
			boundary along with the limit of the boundary at the expiry.
			The position of optimal shouting boundary at expiry, i.e. the boundary of set of positive values for call shout
			option is the same as for put shout option:
			$$
				\partial Z^{+shout}_T=X=S^{*shout}_T.
			$$
			This result can be also found in Alobaidi et al. (2011). However, the value of this limit was derived
			only by argumentation and without any mathematical formulation.
			
			\smallskip
			Now, we present the derivation according to the Theorem~\ref{st_p}. The pay-off function of call and put shout
			option is
			$$
				\Omega^{call,shout}(t,S;X)=\left\{\begin{array}{ll}0&\textrm{for }S\leq X,\\
											S-X+C_{eu}(t,S;S)&\textrm{for }S>X
											\end{array}\right.
			$$
			and
			$$
				\Omega^{put,shout}(t,S;X)=\left\{\begin{array}{ll}X-S+P_{eu}(t,S;S)&\textrm{for }S<X,\\
											0&\textrm{for }S\geq X,
											\end{array}\right.
			$$
			respectively. The functions $C_{eu}$ and $P_{eu}$ are defined by (\ref{call_value}) and
			(\ref{put_value}), respectively.
			
			Notice, that the underlying $S$ is under the same measure as for the vanilla option, thus we have the numeraire
			$\mathcal{N}_t=e^{rt}$. In this case, we use the idea from (\ref{bonus_deriv})
			to determine the bonus function $f_{b}$:
			$$
				f^{call,shout}_{b}(T,S)=\left\{\begin{array}{ll}0&\textrm{for }S<X,\\
														\infty&\textrm{for }S\geq X
											\end{array}\right.
				\quad\textrm{and}\quad
				f^{put,shout}_{b}(T,S)=\left\{\begin{array}{ll}\infty&\textrm{for }S\leq X,\\
														0&\textrm{for }S>X.
											\end{array}\right.
			$$
			
	\subsection{British vanilla options}
	
			The British vanilla option is financial derivative hedging the real trend of the underlying asset.
			This feature allows its holder to exercise the option prior to the expiry $T$ and receive the
			best prediction of the pay-off according to the real trend of underlying restricted to the contract drift $\mu_c$.
			
			The position of the early exercise boundary for British plain vanilla call and put option is equal to
			the boundary of set of positive values of $f^{GB,call}_{b}$ and $f^{GB,put}_{b}$, i.e.
			$$
				\partial Z^{+GB,call}_T=\max\left[X,\frac{r}{q+\mu_c}X\right]=S^{*GB,call}_T
			$$
			and
			$$
				\partial Z^{+GB,call}_T=\min\left[X,\frac{r}{q+\mu_c}X\right]=S^{*GB,call}_T,
			$$
			respectively. This result is consistent with the calculation presented in Peskir and Samee (2008a,b).
			However, the derivation presented in this paper is much shorter and straightforward.
			
			\smallskip
			Now, we present the derivation according to the method presented in this paper.
			The pay-off functions of call and put British vanilla option are
			$$
					\Omega^{GB,call}=e^{\mu_c (T-t)}S{\bf\Phi}\left(d_t^{\mu_c}\right)
							-X{\bf\Phi}\left(d_t^{\mu_c}-\sigma\sqrt{T-t}\right)
			$$
			and
			$$
					\Omega^{GB,put}=X{\bf\Phi}\left(-d_t^{\mu_c}+\sigma\sqrt{T-t}\right)
									-e^{\mu_c (T-t)}S{\bf\Phi}\left(-d_t^{\mu_c}\right),
			$$
			respectively. The function $\Phi(\ \cdot\ )$ is standard normal cumulative distribution function and
			the expression $d_t^{\mu_c}=\frac{\ln{\frac{S}{X}}+(\mu_c+\frac12\sigma^2)(T-t)}{\sigma\sqrt{T-t}}$.
			
			Notice again, that the underlying $S$ is under the same measure as for the vanilla option, thus we have the
			numeraire $\mathcal{N}_t=e^{rt}$. We can use the idea from (\ref{bonus_deriv})
			to determine the bonus function $f_{b}$:
			$$
				f^{GB,call}_{b}(T,S)=\left\{\begin{array}{ll}0&\textrm{for }S<X,\\
														(q+\mu_c) S - r X&\textrm{for }S\geq X
											\end{array}\right.
			$$
			and
			$$
				f^{GB,put}_{b}(T,S)=\left\{\begin{array}{ll}r X - (q+\mu_c) S&\textrm{for }S\leq X,\\
														0&\textrm{for }S>X.
											\end{array}\right.
			$$
	
\section{Verification by numerical method}

	In this section, we compare and verify analytic values of the limit of early exercise boundary
	with values calculated by the PSOR (projected successive over relaxation) method introduced in
	Elliot and Ockendom (1982). For further details on the PSOR method see Kwok (2008) or Wilmott et al. (1995).
	Values for plain vanilla options, Asian options and lookback options were derived
	in other sources. We present only the results of comparison for condor spread as a case with several different
	values of the limit at the expiry (results for all of the other examples are very similar).
	
	In Table~\ref{condor_tab}, we present values of the limit of early exercise boundary at expiry
	calculated by the PSOR method with 250 time and 40000 space steps. The space domain was bounded
	to the interval $[-1.5,1.5]$. The SOR was calculated with parameter $\omega=1.4$ and tolerance $\varepsilon=10^{-14}$.
	The expiration time has to be chosen close to zero, we set the value $T=10^{-8}$ in the calculation.
		
	Each numerical value tends to analytic value calculated by the formulae presented in section \ref{calc}.
	The numerical results are improving with increasing density of the time--space mesh and decreasing tolerance
	(as expected). The relative error of results with the highest precision is of the order $10^{-4}$.
		
		\begin{table}[ht]
		\begin{center}
			\caption{\label{condor_tab}Comparison of analytic and numerical values of the limit of early exercise boundary for condor spread.}
				{\begin{tabular}{r|r|r|r|r|r|r||r|r|r}
			\multicolumn{1}{c|}{$r$}&\multicolumn{1}{|c|}{$q$}&\multicolumn{1}{|c|}{$\sigma$}&\multicolumn{1}{|c|}{$X_1$}&\multicolumn{1}{|c|}{$X_2$}&\multicolumn{1}{|c|}{$X_3$}&\multicolumn{1}{|c||}{$X_4$}&\multicolumn{1}{|c|}{$S^*_{theor}$}&\multicolumn{1}{|c|}{$S^*_{calc}$}&\multicolumn{1}{|c}{error}\\
			\hline\hline
	$3\%$	&	$2\%$	&	$30\%$	&	1	&	3	&	4	&	5	&	\begin{tabular}{r}1.5\end{tabular}	&	\begin{tabular}{r}1.50002\end{tabular}	&	\begin{tabular}{r}$ 0.001\%$\end{tabular}	\\
	\hline\hline
	$2\%$	&	$3\%$	&	$30\%$	&	1	&	3	&	4	&	5	&	\begin{tabular}{r}1\\4\\5\end{tabular}	&	\begin{tabular}{r}1.00033\\4.00074\\5.00159\end{tabular}	&	\begin{tabular}{r}$0.033\%$\\$0.019\%$\\$0.032\%$\end{tabular}	\\
	\hline\hline
	$3\%$	&	$2\%$	&	$30\%$	&	1	&	2	&	3	&	4.5	&	\begin{tabular}{r}1.5\\4.5\end{tabular}	&	\begin{tabular}{r}1.50006\\4.49844\end{tabular}	&	\begin{tabular}{r}$  0.004\%$\\$-0.035\%$\end{tabular}		
				\end{tabular}}
		\end{center}
		\end{table}

\section{Conclusions}

	In this paper, we presented a unified method for calculation of a limit of the early exercise boundary at expiry.
	The method is applicable for American style of a wide range of financial derivatives. We calculated and compared the
	analytic value of early exercise boundary limit for plain vanilla options as well as for Asian and lookback options.
	The results coincides with the well known values that can be found in Albanese and Campolieti (2006), Alobaidi and Mallier (2006),
	Bokes and {\v S}ev{\v c}ovi{\v c} (2011), Chiarella and Ziogas (2005), Dai and Kwok (2006), Detemple (2006), Kwok (2008),
	{\v S}ev{\v c}ovi{\v c} (2008), Wilmott et al. (1995), Wu et al. (1999)
	and in many other sources. Moreover, we also calculated the analytic value of the limit for
	shout option and American style of exponentially weighted floating strike Asian option with general average, British vanilla
	option and option strategies (represented by condor spread). We verified values by their comparison with values calculated
	by the PSOR method.
	Presented straightforward method is simple and unifies the approach to methods so far used for the calculation
	of limit of exercise boundary.

\section*{Acknowledgments}
	
	This research was supported by VEGA 1/0381/09 and APVV SK-BG-0034-08 grants.

\end{document}